\documentclass[aps,prc,preprint,superscriptaddress,showkeys,showpacs]{revtex4}

\usepackage{graphicx}

\begin{document}

\title{Boson stars with repulsive selfinteractions}

\author{Pratik Agnihotri}

\affiliation{Indian Institute of Technology, Kanpur, India}

\affiliation{Institut f\"ur Theoretische Physik, 
J. W. Goethe Universit\"at, Max von Laue-Stra\ss{}e~1, 
D-60438 Frankfurt am Main, Germany}
 
\author{J\"urgen Schaffner-Bielich}

\affiliation{Institut f\"ur Theoretische Physik,
Ruprecht-Karls-Universit\"at, Philosophenweg 16,
D-69120 Heidelberg, Germany}

\author{Igor N. Mishustin}

\affiliation{Frankfurt Institute for Advanced Studies, 
J. W. Goethe Universit\"at, Ruth-Moufang-Stra\ss{}e~1, 
D-60438 Frankfurt am Main, Germany}

\affiliation{The Kurchatov Institute, Russian Research Center, 
123182 Moscow, Russia}

\date{\today}

\begin{abstract}
  The properties of compact stars made of massive bosons with a
  repulsive selfinteraction mediated by vector mesons are studied
  within the mean-field approximation and general relativity. We
  demonstrate that there exists a scaling property for the mass-radius
  curve for arbitrary boson masses and interaction strengths which
  results in an universal mass-radius relation. The radius remains
  nearly constant for a wide range of compact star masses.  The
  maximum stable mass and radius of boson stars are determined by the
  interaction strength and scale with the Landau mass and radius.
  Both, the maximum mass and the corresponding radius increase
  linearly with the interaction strength so that they can be radically
  different compared to the other families of boson stars where interactions
  are ignored.
\end{abstract}



\maketitle


\section{Introduction}


White dwarfs, neutron and quark stars, collectively dubbed
compact stars, are the final result of stellar evolution.  White
dwarfs are stabilized by the Fermi degeneracy pressure. There exists
an upper limit for the mass of a white dwarf, the Chandrasekhar mass,
which is about 1.4 times the mass of the sun
\cite{1931ApJ....74...81C}.  Beyond this limit the white dwarf is
unstable against gravitational collapse. Neutron stars are stable
mainly due to the repulsive nature of the interactions between
nucleons. Therefore, the precise value of the maximum mass for a
neutron star is less certain, but is presumably close to the
predictions based on the Landau consideration \cite{Landau32}. As
shown in Ref.~\cite{2006PhRvD..74f3003N}, Landau's argument can be
extended to a general compact star made of fermions, a fermion star,
with arbitrary fermion mass and interaction strength.

In the following we are studying compact stars made of bosons. Unlike
fermion stars, boson stars have no observational evidence,
yet. Besides this the existence of any stable scalar particle has
never been experimentally verified. Wheeler \cite{1955PhRv...97..511W}
introduced a gravitational electromagnetic entity called a geon. The
gravitational attraction of its own field energy confines the geon in
a certain region. Later Kaup \cite{1968PhRv..172.1331K} has solved the
Klein-Gordon Einstein equations for scalar fields and found a new
class of solutions for gravitating objects. These boson stars are
stable with respect to spherically-symmetric gravitational
collapse. Ruffini and Bonazzola \cite{1969PhRv..187.1767R}
demonstrated that boson stars describe a family of self-gravitating
scalar field configurations within general relativity.  In
ref.~\cite{1984ZPhC...26..241T} Takasugi and Yoshimura calculated
boson stars within an approach similar to the one conventionally
adopted for neutron stars by solving the Tolman-Oppenheimer-Volkoff
(TOV) equation \cite{Tolman34,Tolman39,OV39} with a separate equation
of state describing the properties of matter.  Boson stars with
selfinteractions have been also considered, in particular as
candidates for dark matter \cite{Colpi:1986ye}. For reviews on boson
stars we refer to
\cite{Lee:1991ax,1992PhR...220..163J,2003CQGra..20R.301S}.

In the following, we consider boson stars as localized,
gravitationally bound objects made of self-interacting bosons at zero
temperature. The interactions between the bosons is described by
vector meson exchange in the relativistic mean-field approximation.
The resulting equation of state is used as input to solve the TOV
equation for boson stars, similar to the approach of
ref.~\cite{1984ZPhC...26..241T} but with an equation of state based on a field-theoretical approach.  We demonstrate that there are
scaling relations for the mass-radius curve. In particular, we show
that the maximum mass is controlled by the interaction strength and
the Landau mass, not by the boson mass. We compare our results to
previous works and to the case of fermion stars with
self-interactions.


\section{Scaling relations for compact stars}


We assume a spherically-symmetric and static configuration where the
energy-momentum tensor is that of a perfect fluid at rest. Then the star
structure can be obtained by solving the Tolman-Oppenheimer-Volkoff
(TOV) equations, which can be conveniently written as
\begin{equation}
\frac{dp}{dr}=-\frac{GM\rho }{r^{2}}\left(1+\frac{p}{\rho }\right)
\left(1+\frac{4\pi r^{3}p}{M}\right)
\left(1-\frac{2GM}{r}\right)^{-1} \label{c}
\end{equation}
with
\begin{equation}
\frac{dM}{dr}=4\pi r^{2}\rho \label{d} \quad .
\end{equation}
Additionally, one needs an equation of state, $p(\rho)$, which
describes the microscopic properties of the stellar matter. These
coupled differential equations for the pressure $p(r)$ and the mass
profile $M(r)$ are integrated from $r=0$ with some central value
$p_{0}$ to a point where the pressure vanishes $p(R)=0$ which defines
the radius $R$ and the total mass $M(R)$ of the boson star.

The TOV equations have a similar scaling behavior as the one for
Newtonian hydrostatic equilibrium (see e.g.\
\cite{2006PhRvD..74f3003N}). Just the gravitational constant $G$ and
the boson mass $m_{b}$ are used to rewrite the TOV equations together 
with the equation of state in dimensionless form.
Solving the dimensionless TOV equation for a
certain class of equations of state allows for deriving general
solutions by just rescaling the results by appropriate dimensionfull
parameters.

The first relativistic correction factor in eq.~(\ref{c}), i.e.\
$(1+p/\rho)$, can be scaled by choosing $p^{\prime}=p/\rho_{o}$ and
$\rho^{\prime}=\rho/\rho_{o}$. Here $\rho_{o}$ is a common factor with
dimension of mass to the fourth power. For the other two relativistic
factors in eq.~(\ref{c}) we introduce the dimensionless mass
$M^{\prime}=M/a$ and the dimensionless radius $r^{\prime }=r/b$. For a
dimensionless expression one has to set
\begin{equation}
\frac{b^{3}\rho_{o}}{a}=1 \qquad \mbox{and} \qquad
\frac{a}{M_{P}^{2}b}=1 \label{i}
\end{equation}
that leads to the following relations
\begin{equation}
a=\frac{M_{P}^{3}}{\sqrt{\rho_{o}}} \qquad \mbox{and} \qquad 
b=\frac{M_{P}}{\sqrt{\rho_{o}}} \label{j}
\end{equation}
where $M_P=\sqrt{\hbar c/G}$ is the Planck mass.  Note that the
Newtonian terms do not give any additional constraint.  Choosing
$\rho_{o} = m_b^4,$ the rescaling factors are $a=M_{P}^{3}/m_b^2$ and
$b=M_{P}/m_b^2$, which coincide with the expressions of the maximum
mass and the radius for compact stars introduced by Landau
\cite{Landau32},
\begin{equation}
M_{L}=\frac{M_{P}^{3}}{m_{b}^{2}} \qquad \mbox{and} \qquad
R_{L}=\frac{M_{P}}{m_{b}^{2}} \label{f}
\end{equation}
We note in passing that for the MIT bag equation of state these scaling
factors are $\rho_{o}=B$ so that the maximum mass and the
corresponding radius scale as $B^{-1/2}$, a well known result for quark stars
\cite{Witten:1984rs}.


\section{Meson exchange model for interacting bosons}


We describe the interactions between scalar bosons by the exchange of
vector mesons. For a scalar field $\phi$ and a vector field $V_\mu$ the
Lagrangian reads
\begin{equation}
{\cal L} = {\cal D}_\mu^* \phi^* {\cal D}^\mu \phi - {m_b}^2 \phi^*\phi
- \frac{1}{4} V_{\mu\nu} V^{\mu\nu} + \frac{1}{2}m_v^2 V_\mu V^\mu
\end{equation}
with $V_{\mu\nu}=\partial_\mu V_\nu -\partial_\nu V_\mu$.
The boson field is coupled to the vector field by a minimal coupling scheme
\begin{equation}
{\cal D}_\mu = \partial_\mu + i g_{v \phi} V_\mu \quad .
\end{equation}
where $g_{v\phi}$ is the $\phi$-$V$ coupling strength. 
Note, that the vector field has a quadratic coupling term to the
scalar field in the Lagrangian which ensures that the vector field is
coupled to a conserved current (see below). We treat the vector field
as a classical field. In static bulk matter the spatial components of the
vector field vanish and the equation of motion for the scalar field
reads
\begin{equation}
\left[{\cal D}^{*}_{\mu} {\cal D}^\mu
 + {m_b}^2 \right] \phi(x) = 0 \quad .
\label{eq:Lagr}
\end{equation}
In the mean-field approximation, after expanding into plane waves, we
obtain for the lowest energy mode $k=0$:
\begin{equation}
\omega_\phi = m_b + g_{v \phi} V_0 
\label{eq:disp}
\end{equation}
Note that the vector interaction between the scalar particles is
repulsive which ensures the overall stability of selfinteracting boson
matter. The vector field is determined from the equation
\begin{equation}
m_v^2 V_0 =  2 g_{v \phi} \left(\omega_\phi - g_{v\phi} V_0\right) \phi^*\phi = 
2 g_{v \phi} m_b \phi^* \phi
\end{equation}
where we have used the dispersion relation for the $\phi$ field
eq.~(\ref{eq:disp}).  The conserved current for the scalar field can
be obtained from the Lagrangian (\ref{eq:Lagr})
\begin{eqnarray}
J_\mu &=& i \left(\phi^* \frac{\partial \cal L}{\partial^\mu \phi^*}
- \frac{\partial \cal L}{\partial^\mu \phi} \phi \right) \cr
&=& \phi^* i\partial_\mu \phi -(i\partial_\mu \phi^*)\phi
+ 2g_{v \phi} V_\mu\phi^* \phi \quad .
\end{eqnarray}
The number density of bosons 
\begin{equation}
n_b = J_0 = 2\left(\omega_\phi - g_{v \phi} V_0\right) \phi^* \phi  
= 2 m_b \phi^* \phi
\end{equation}
is just the source term for the vector field. The total energy density
of the boson matter can be determined from the energy-momentum tensor
\begin{equation}
\rho =  2 m_b^2 \phi^*\phi + \frac{1}{2}m_v^2 V_0^2
= m_b n_b + \frac{g_{v\phi}^2}{2 m_v^2}  n_b^2
\end{equation}
where the equation of motion for the vector field has been used.
The pressure is given just by the vector field contribution
\begin{equation}
p = \frac{1}{2}m_v^2 V_0^2 = \frac{g_{v\phi}^2}{2 m_v^2} n_b^2
\end{equation}
Note, that these expressions are thermodynamically consistent as can be
checked by using the thermodynamic relation
\begin{equation}
p = n_{b}^{2}\frac{d\left(\rho/n_b\right)}{dn_{b}} \label{m}
\end{equation}
The form of the interaction is actually similar to the one used for
interacting fermions and the corresponding Fermi stars in
\cite{2006PhRvD..74f3003N}. For the rescaled TOV equations we introduce
the dimensionless interaction parameter $y=m_b/m_I$, where
$m_I=\sqrt{2}m_v/g_{v\phi}$, and set $\rho^{\prime}=\rho/m_{b}^{4}$ and
$p^{\prime }=p/m_{b}^{4}$. The equation of state for interacting boson
matter can be summarized to be of the simple form
\begin{equation}
p' = y^2 {n'_{b}}^2
\qquad \mbox{and} \qquad
\rho' = n'_{b}+ y^2 {n'_{b}}^2
\end{equation}
with the dimensionless number density $n'_b = n_b/m_b^3$.  It is
possible to represent the equation of state in a polytropic form
$p=\rho^\gamma$ for certain limits.  For low densities, one approaches
$p\propto\rho^2$, a polytrope with $\gamma = 2$.  For high densities,
one has an equation of state of the form $p=\rho$ with $\gamma =1$,
which is the stiffest possible equation of state first discussed by
Zeldovich \cite{Zeldovich61}. The switch between those two limiting cases
is controlled by the interaction strength $y$. The larger $y$, the
lower is the energy density to approach the causal limit $p=\rho$.

\begin{figure}
\includegraphics[width=0.7\textwidth]{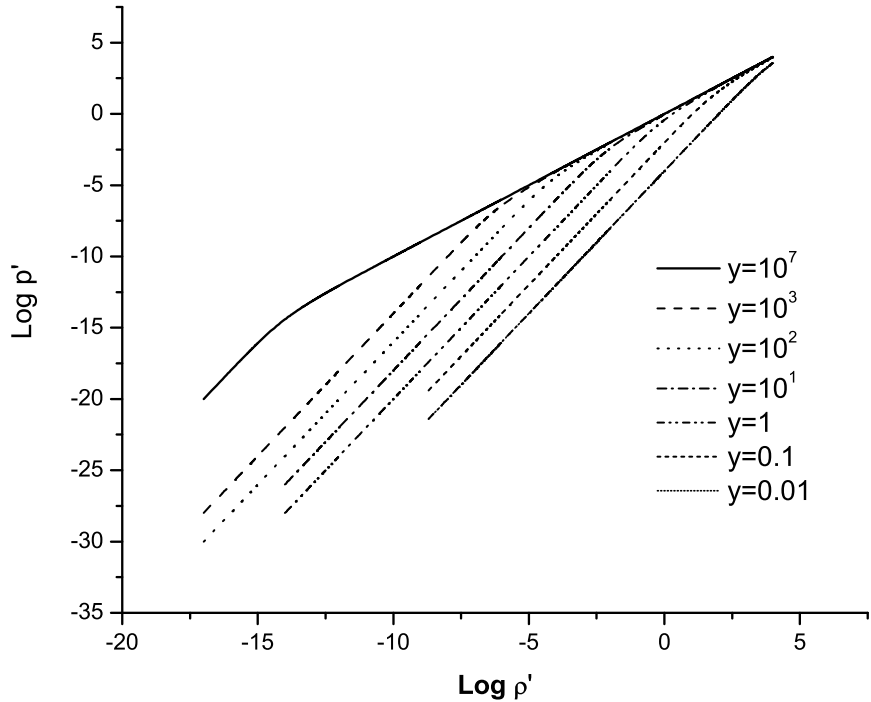}
\caption{\label{fig:eos} Double logarithmic plot of the dimensionless
  pressure versus the dimensionless energy density for different
  interaction strengths $y$.}
\end{figure}

Fig.~\ref{fig:eos} depicts the dimensionless pressure versus the
dimensionless energy density.  The values of the interaction strength
$y$ are chosen between $ 10^{-2}$ to $10^{7}$. There are two different
slopes for small and large values of $\rho ^{\prime }$ corresponding
to the above mentioned limits. The point where the slope changes shifts to lower
densities with increasing interaction strength $y$, but for low densities the slope is
$\gamma=2$. At high $\rho
^{\prime }$ all curves merge to the limiting curve with a slope of
$\gamma = 1$.


\section{Scaling relation for boson stars with selfinteraction}

\begin{figure}
\includegraphics[width=0.7\textwidth]{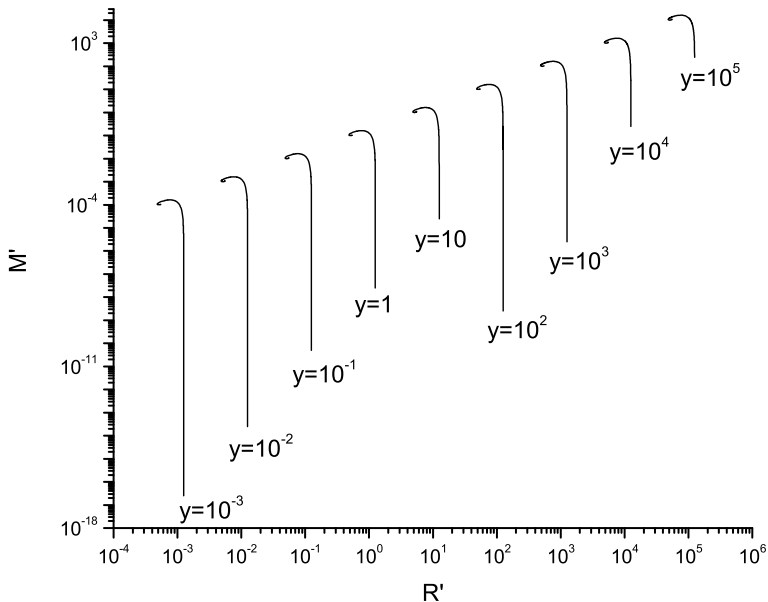}
\caption{\label{fig:mr} The dimensionless mass $M^{\prime }$ is
  plotted versus the dimensionless radius $R^{\prime }$ for different
  interaction strengths $y$ ranging from $10^{-3}$ to $10^{5}$. Note
  that each mass-radius curve terminates in a spiral at the left end,
  which is not visible in the double logarithmic plot, but can be seen
  in the linear plot of Fig.~\ref{fig:massradius}.}
\end{figure}

We use the dimensionless equation
of state to solve the dimensionless TOV equations.  The equation of
state depends only on the interaction strength $y$. Fig.~\ref{fig:mr}
shows the double logarithmic plot of the dimensionless mass $M^{\prime}$
versus the dimensionless radius $R^{\prime}$ for different interaction
strengths $y$ ranging from $10^{-3}$ to $10^{5}$.  One observes that
each mass-radius curve contains a constant radius part over a wide range of
masses. Also, the curves are very similar and seem to be just shifted
to larger masses and radii with increasing interaction strength.

This interesting behavior can be explained by considering the equation of
state described by a polytrope $p\sim \rho^\gamma$ with $\gamma=2$.
In general, the solution to the Lane-Emden equation, see
e.g.\ \cite{Weinberg72}, results in a mass-radius relation of the form
$M^{\prime }\propto\rho_{c}^{(3\gamma-4)/2}$ and
$R^{\prime}\propto\rho_{c}^{(\gamma -2)/2}$, where $\rho_{c}$ is the
central energy density. Hence, at low densities and large radius
$R\gg 2GM$, where effects from general relativity can be ignored, the
mass of the star increases linearly with $\rho _{c}$ while the radius
$R^{\prime}$ remains constant for $\gamma=2$ that explains the
peculiar form of the mass-radius curves.

\begin{figure}
\includegraphics[width=0.7\textwidth]{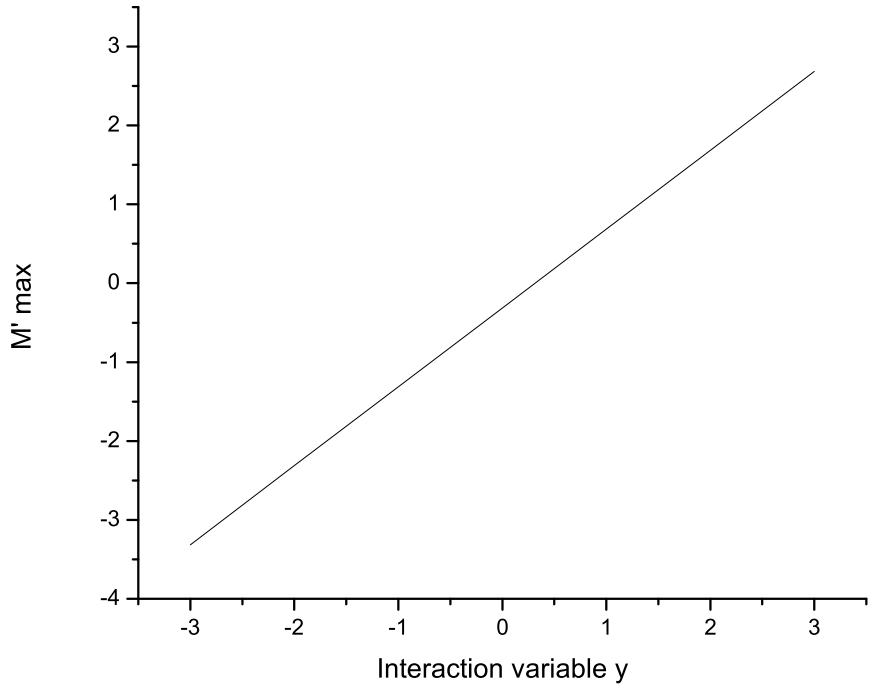}
\caption{\label{fig:maxmass} The dimensionless maximum mass
  $M_{\max}^{\prime}$ is plotted as a function of interaction strength
  $y$.}
\end{figure}

There exist another interesting feature of the mass-radius curves
which reflects the scaling properties of the equation of state and the
TOV equations. To illustrate this we plot in Fig.~\ref{fig:maxmass}
the dimensionless maximum mass $M_{\max}^{\prime}$ as a function of
the interaction strength $y$. It is interesting to see that the maximum mass
$M_{\max}^{\prime}$ scales linearly with the interaction strength $y$.

\begin{figure}
\includegraphics[width=0.7\textwidth]{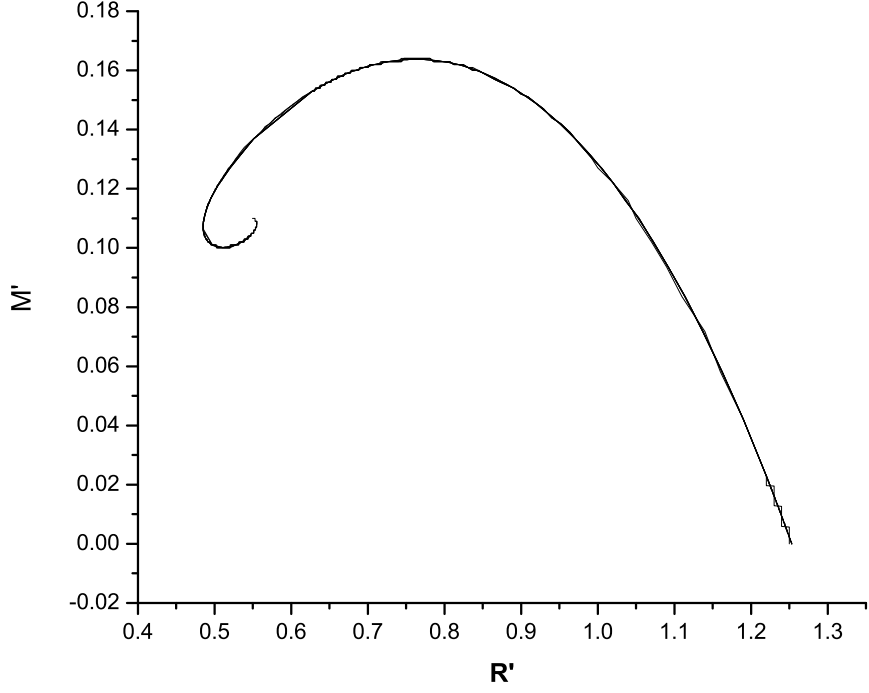}
\caption{\label{fig:mrscaled} The dimensionless mass $M^{\prime}$ is plotted
  versus the dimensionless radius $R'$ dividing both by the interaction
  strength $y$. The result is an universal mass-radius relation.}
\label{fig:massradius}
\end{figure}

Therefore, one can conclude that by proper rescaling all mass-radius
curves can be reduced to one universal mass-radius curve.  Indeed,
dividing the dimensionless mass $M^{\prime }$ and the corresponding
radius $R'$ by the interaction strength $y$ results in an unique
mass-radius relation as depicted in Fig.~\ref{fig:mrscaled}. This
graph looks rather similar to the mass-radius curve of a strongly
interacting fermion star \cite{2006PhRvD..74f3003N}. There the maximum
mass is constant for weak interactions ($y\ll 1$) and increases
linearly in $y$ for strong interactions ($y\gg 1$).  Note that the
part of the curve to the left of the maximum mass represents unstable
configurations, only the star configurations at the maximum and to the
right of it can exist. The existence of a maximum mass is determined
by the change of the equation of state from a polytrope with
$\gamma=2$ to one with $\gamma=1$. The latter value is lower than the
critical value $\gamma_c=4/3$ for stable compact stars (effects of
general relativity will even increase this value slightly).

\begin{figure}
\includegraphics[width=0.7\textwidth]{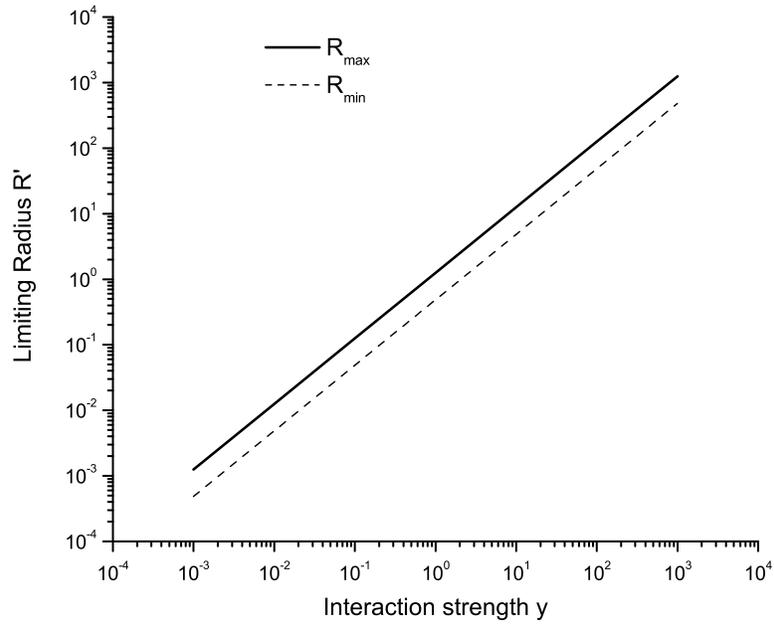}
\caption{\label{fig:radii} Plot of the two limiting radii
  $R_{\max}^{\prime}$ and $R_{\min }^{\prime}$ for boson stars as a
  function of interaction strength $y$.}
\end{figure}

Fig.~\ref{fig:radii} shows the two limiting radii for interacting
boson stars as a function of the interaction strength $y$.  Here,
$R_{\max}^{\prime}$ and $R_{\min}^{\prime}$ denote the maximum and
minimum radius for boson stars, respectively. $R'_{\min}$ stands for
radius corresponding to the maximum mass configuration, while
$R'_{\max}$ is the radius of stars with masses much smaller than the
maximum mass. Both, the maximum and the minimum radius vary linearly
with the interaction strength $y$ and the difference between the two
radii is rather small, by a factor of about 0.61 independent of the
interaction strength.

\begin{figure}
\includegraphics[width=0.7\textwidth]{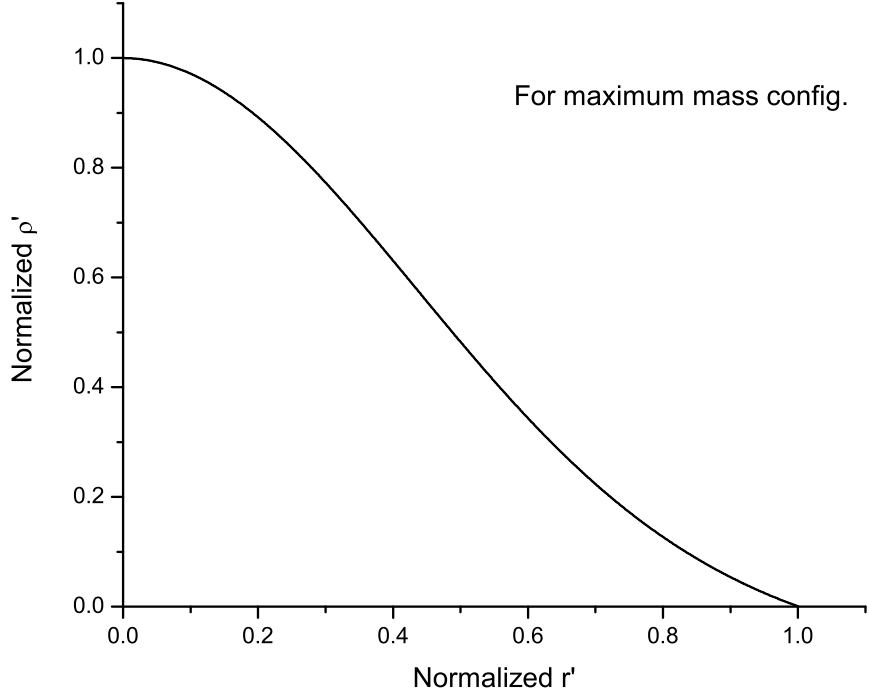}
\caption{\label{fig:profile} The variation of the normalized density
  $\rho'(r)/\rho'(0)$ with the normalized radius $r'/R'$ for different
  interaction strengths showing that there is only one universal curve.}
\end{figure}

Fig.~\ref{fig:profile} shows the normalized density profile of
$\rho'(r)/\rho'(0)$ over the normalized radius $r'/R'$ for different
interaction strengths calculated for the maximum mass configuration.
Again, there appears an universal curve independent of the interaction
strength and the mass of the boson.  The rate of the decrease of the
density with the radius is then the same for all interaction strengths
$y$. The density profile shows a small plateau in the core region up
to a radius of about $r\sim 0.1R$ followed by a nearly linear decrease
up to the surface of the boson star.

The scaling behavior observed for interacting boson stars follows in
straightforward way from our discussion above on the scaling features of
the TOV equations.  The maximum mass configuration is determined by
the equation of state at high central densities, where $p=\rho$. The
pressure and the energy density, $p$ and $\rho$, depend then on the
interaction strength as $y^2$.  Rescaling both, the energy density and
the pressure, by the factor $m_b^4/y^2$ gives the modified Landau mass
and Landau radius
\begin{equation}
  M^b_{L}=y\cdot\frac{M_{P}^{3}}{m_{b}^{2}} \qquad \mbox{and} \qquad
  R^b_{L}=y\cdot\frac{M_{P}}{m_{b}^{2}} 
\end{equation}
for compact stars with interacting bosons. The maximum mass and the
corresponding radius have to increase linearly with the interaction
strength $y$. This is in complete agreement with the results of
Ref.~\cite{Colpi:1986ye} where the selfinteraction term of the
form $\lambda\phi^4$ was used to describe the interactions between bosons.
Also for this type of interactions, the maximum mass and the
corresponding radius were found to scale with the interaction strength
and the Landau mass as $M_{\max}\propto \lambda^{1/2} M_L$. The
relation for the maximum mass is compatible with our findings by
realizing that the dimensionless coupling constant $\lambda$ can be
associated with our interaction strength $y^2$. In addition to the
case of scalar selfinteraction, we find that for vector interactions
the scaling property is even more general as the whole mass-radius
curve can be described by an universal curve when using the modified
Landau mass and radius. We note, that compact stars made of fermions
with vector interactions \cite{2006PhRvD..74f3003N} reveal the same
scaling feature of the mass-radius curve for large interaction
strengths.

The maximum mass of boson stars as obtained from numerical
calculations is
\begin{equation}
M_{\max } \approx 0.164 y \cdot \frac{M_{P}^{3}}{m_{b}^{2}} \label{w}
\end{equation}
and the two limiting radii of boson stars are given by
the expressions
\begin{equation}
R_{\max }\approx 1.252 y\cdot \frac{M_{P}}{m_{b}^{2}} \qquad \mbox{and} \qquad
R_{\min }\approx 0.763 y\cdot \frac{M_{P}}{m_{b}^{2}} \label{y}
\end{equation}
The above relations can be used to calculate the maximum mass and the
maximum and minimum radii of boson stars for arbitrary interaction
strength $y$ and boson masses $m_b$. The values for the Landau mass
and radius are $M_L=1.632 M_{\odot }$ and $R_L=2.410$~km,
respectively, for a boson mass of $m_b=1$~GeV.  One recovers the same scaling
relations as for the noninteracting case, see e.g.\
Ref.~\cite{1984ZPhC...26..241T}, by setting the interaction scale equal to
the Planck mass, $m_{I}= M_{P}$,
\begin{equation}
M_{\max }\propto \frac{M_{P}^{2}}{m_{b}} 
\qquad \mbox{and} \qquad
R_{\min } \propto \frac{1}{m_{b}}
\qquad ,
\end{equation}
which are orders of magnitude smaller than for the case of realistic
interactions. We note that our numerical prefactors are different from the ones of Takasugi and Yoshimura \cite{1984ZPhC...26..241T} while the scaling with the boson mass is the same. These authors adopt a different equation of state, where the pressure has the form as for degenerate configurations, e.g.\ in the low-density limit they recover that $p\propto \rho^{5/3}$. In our case the pressure is determined by interactions only and is proportional to the density squared.

\begin{table}
\begin{ruledtabular} 
\begin{tabular}[t]{cccccccc}
boson mass & 100 GeV & 1 GeV & 1 MeV & 1 keV & 1 eV & $10^{-5}$
eV & interaction \\
\hline
$M_{\max }(M_{\odot })$ & $10^{-22}$ & $10^{-20}$ & $10^{-17}$ & $10^{-14}$ & $10^{-11}$ 
& $10^{-6}$ & $m_I=M_P$ \\
$R$ (km) & $10^{-21}$  & $10^{-19}$  & $10^{-16}$  & $10^{-13}$  & $10^{-10}$
 & $10^{-5}$ & (free case) \\
\hline
$M_{\max }(M_{\odot })$ & $10^{-5}$ & $0.1$ & $10^{5}$ & $10^{11}$ & $10^{17}
$ & $10^{27}$ & $y=1$ \\
$R$ (km) & $10^{-4}$  & $1$  & $10^{6}$  & $10^{12}$  & $10^{18}$
 & $10^{28}$ & (Landau limit) \\
\hline
$M_{\max }(M_{\odot })$ & $10^{-5}$ & $10^{-3}$ & $1$ & $10^{3}$ & $10^{6}$
& $10^{11}$ & $m_I = 100$ GeV \\
$R$ (km) & $10^{-4}$  & $10^{-2}$  & $10$  & $10^{4}$  & $10^{7}$
 & $10^{12}$ & (weak scale) \\
\hline
$M_{\max }(M_{\odot })$ & $10^{-2}$ & $1$ & $10^{3}$ & $10^{6}$ & $10^{9}$ & $
10^{14}$ & $m_I= 100$ MeV \\
$R$ (km) & $0.1$  & $10$  & $10^{4}$  & $10^{7}$  & $10^{10}$ 
& $10^{15}$  & (QCD scale)
\end{tabular}
\end{ruledtabular}
\caption{
  Order of magnitude scales of the maximum mass and the characteristic
  radius of compact stars made of different boson masses and
  interaction strengths. The first set corresponds to the free case by
  setting $y=m_b/M_P$ (the boson stars are just bound by gravity), the
  second gives the Landau mass and radius by setting $y=1$, which
  holds for boson and fermion stars. The third and last set lists the
  values for interaction mass scales of the standard model weak and
  strong interactions, i.e.\ of 100~GeV and 100~MeV, respectively.} 
\label{tab:examples}
\end{table}

Table~\ref{tab:examples} gives the maximum mass and the corresponding
radius for four different cases of the interaction parameter. By
setting $m_I=M_P$ or $y=m_b/M_P$ one recovers the case for ordinary
boson stars with free bosons (see above) which are just
gravitationally bound. The case $y=1$ gives the Landau mass and radius
of compact stars, which is nearly the same for boson stars and fermion
stars when including interactions. Finally, we consider the case of
interactions mediated by the weak interaction scale of about
$m_{I}=100$~GeV and the QCD scale of about $m_{I}=100$~MeV. For the
boson masses we choose the range from the electroweak scale to a
typical mass of axions, $\sim 10^{-5}$~eV. We want to emphasize the
following features of these calculations.  For the free case $m_I=M_P$
one only reaches astrophysically interesting scales for boson masses
of less than $10^{-5}$~eV.  Maximum masses close to the ones for
neutron stars, $M\sim 1M_\odot$, can be reached by boson stars for
boson masses of around 1~GeV (Landau case), 1~MeV for bosons with weak
interactions, and 1~GeV for strong interactions. The mass range of
observed supermassive black holes, $M=10^6$ to $10^9 M_\odot$, is
found for boson masses between 1~keV and 1~MeV for the Landau case,
1~eV and below for weak interactions, and 1~eV to 1~keV for strong
interactions.  It is observed that the inclusion of interactions
results in a wide range of possible masses and radii for boson stars,
covering scales as small as a fraction of a solar mass and below a
kilometer to scales of supermassive black holes.  It is interesting to
note, that a boson with a mass of 100~GeV and with QCD-type
interaction strengths gives star configurations with masses and radii
as a neutron star. Surprisingly, for a boson star made of axions
($m_b\sim 10^{-5}$~eV) and selfinteractions on the scale of
$10^{12}$~GeV, one obtains a mass of about $30M_\odot$ with a radius
of 200~km, i.e.\ the mass of compact objects found in binary systems
which are attributed to light black holes. These values are orders of
magnitude different compared to the case of boson stars with
noninteracting axions, see also ref.~\cite{1984ZPhC...26..241T} and
table~\ref{tab:examples}.


\section{Summary}


We have constructed an equation of state for a system of massive
bosons interacting by the exchange of vector mesons. By solving the
TOV equations for such boson stars, we have demonstrated that there exists
a universal mass-radius curve independent of the boson mass and the
interaction strength.  The maximum mass and the corresponding radius
of boson stars are scaled with the Landau mass and Landau radius times
the interaction strength. For masses much smaller than the maximum
mass, the radius stays constant and is only slightly larger than the
one for the maximum mass configuration.

The maximum mass and the corresponding radius can be computed with the
simple formulae $M_{\max }=0.164 y\cdot M_{P}^{3}/m_{b}^{2}$ and
$R_{\min}=0.763 y\cdot M_{P}/m_{b}^{2}$ for any given boson mass $m_b$
and interaction strength $y$. The possible masses and radii for boson
stars can therefore cover a wide range and can be similar to the ones
found for astrophysical compact objects, be it neutron stars or black
hole candidates. For example, for a boson with QCD-type interaction
strength and a boson mass of 100~GeV the maximum mass is $M_{\max
}\sim 0.3 M_{\odot}$ with a radius of about 2~km. A boson with a
typical axion-like mass of $10^{-5}$~eV and an interaction scale of
$10^{12}$~GeV will give a maximum mass of the boson star of
$30M_\odot$ with a radius of 200~km. The compactness of boson stars
for the maximum mass configuration is about $R/(2GM)\approx 2.3$ which
is close to the value found for fermion stars $R/(2GM)\approx 2.4$ in
Ref.~\cite{2006PhRvD..74f3003N}. It is interesting that these values
are below the radius of the innermost stable circular orbit of
nonrotating black holes $R/(2GM)=3$. 

Finally, we mention that the full problem addressed here involves solving the Einstein equations with the coupled system of Klein-Gordon and Proca equations which we leave to address as an interesting extension for future work.

\acknowledgments

This work was supported by GSI Darmstadt and by the German Research
Foundation (DFG) within the framework of the excellence initiative
through the Heidelberg Graduate School of Fundamental Physics.
I.M. acknowledges support from the DFG grant 436 RUS 113/957/0-1 and
the Russian grant NS-3004.2008.2.

\bibliography{myreferences,literat,all}

\end{document}